\documentclass[10pt,twocolumn,letterpaper]{article}

\usepackage{cvpr}              

\usepackage{graphicx}
\usepackage{amsmath}
\usepackage{amssymb}
\usepackage{booktabs}
\usepackage{tikz}
\usetikzlibrary{arrows.meta,positioning,shapes.geometric,calc}

\definecolor{cvprblue}{rgb}{0.21,0.49,0.74}
\usepackage[pagebackref,breaklinks,colorlinks,allcolors=cvprblue]{hyperref}

\def\paperID{*****} 
\def\confName{CVPR}
\def\confYear{2026}

\title{TimeLogic Challenge @ CVPR 2026: Strong MLLMs Meet Evidence-Seeking Agents for Temporal-Logic Video Question Answering}

\author{
Zhaoyang Xu$^{1}$, Xusheng He$^{1}$, Wei Liu$^{1}$, Zhenyang Li$^{2}$, Jianlong Wu$^{1}$\\
$^1$Harbin Institute of Technology (Shenzhen) \qquad $^2$HKUST\\
Team: iLearn\_TimeLogic
}

\begin{document}
\maketitle
\begin{abstract}
Temporal-logic video question answering requires a model to reason about \emph{when} actions occur relative to one another---before, after, until, since, overlap, and multi-event chains---rather than merely \emph{what} is present in a video.
Standard vision--language models (VLMs) typically answer such questions in a single pass over a fixed, uniformly sampled set of frames, which is poorly matched to evidence that is often localized to narrow action boundaries or dispersed across several distant events.
We present an evidence-seeking agent that treats temporal-logic VideoQA as active exploration.
The agent follows a Think--Act--Observe loop driven by a multi-granular sampling toolkit (global overview, structured temporal scan, segment skim, and dense focus), where every observation is interleaved with its absolute timestamp so that temporal relations reduce to numerical comparisons on a shared time axis.
Its behavior is shaped by benchmark structure: a lightweight classifier routes each question to a temporal category, each with a tailored policy, iteration depth, and prompt, while sampling budgets adapt to corpus characteristics and clip length.
The resulting training-free system couples Gemini 3.1 Pro with a temporal-reasoning policy and achieves 77.13 AvgAcc on the official TimeLogic test set.
\end{abstract}
    
\section{Introduction}
\label{sec:intro}

Temporal-logic video question answering (VideoQA) asks models to reason about \emph{when} events occur---before, after, overlap, persistence, and multi-event chains---rather than only \emph{what} appears in a video.
TimeLogic~\cite{swetha2025timelogic} makes this capability explicit through natural-language questions grounded in formal temporal operators.
However, standard VLM inference is poorly matched to this setting.
Most systems uniformly sample a fixed frame set and answer in one pass, but the decisive evidence for a temporal relation is often localized to a short window, such as the boundary between adjacent actions.
Once that evidence is missed, the model has no mechanism to recover it or verify a tentative event ordering.
This failure becomes more pronounced for multi-event chains, where the answer depends on localizing several actions on a shared timeline rather than recognizing a single visual state.
The central challenge is thus to allocate a limited visual budget to the moments most likely to decide the temporal relation, especially when relevant events are sparse or visually subtle.

We therefore frame TimeLogic inference as an \emph{active evidence-seeking} problem, in line with recent tool-guided video-agent systems~\cite{lin2026videoseek}.
Our system uses a ReAct-style~\cite{yao2023react} Think--Act--Observe loop with multi-granular, timestamp-interleaved sampling tools.
Instead of committing after one global sample, the agent repeatedly decides where to inspect next, accumulating timestamped observations and reasoning over the full history.
The available tools cover coarse overview sampling, structured temporal scans, segment-level skimming, and dense focus around a candidate boundary.
Because each returned frame is paired with an absolute timestamp, temporal relations can be checked as comparisons on a common time axis.

The policy is guided by the structure of the benchmark.
The released split contains heterogeneous source corpora and many compound three-event categories.
We therefore classify each question into a temporal category and adapt the initial tool, iteration depth, frame budget, and reasoning prompt accordingly.
Long, action-dense clips receive deeper exploration, while shorter clips can often be handled with denser one-pass coverage.

These choices keep the system simple while encouraging targeted verification before answering.
Our contributions are:
\begin{itemize}[leftmargin=*,noitemsep,topsep=0pt,parsep=0pt,partopsep=0pt]
    \item We formulate TimeLogic inference as Think--Act--Observe evidence seeking with multi-granular, timestamp-interleaved sampling.
    \item We design a training-free, question-type-aware policy that adapts tools, prompts, and budgets to benchmark structure.
    \item With Gemini 3.1 Pro~\cite{geminiteam2026gemini31pro}, the final system achieves 77.13 AvgAcc on the official test set.
\end{itemize}

\clearpage
\section{Benchmark Analysis}
\label{sec:dataset}

We analyze the released TimeLogic benchmark structure~\cite{swetha2025timelogic}, focusing on source-corpus characteristics and question templates, to identify where temporal-logic VideoQA is hardest and to motivate the design choices in Section~\ref{sec:method}.
The split contains $3{,}000$ questions over $1{,}850$ unique videos, drawn from four source corpora and posed in two answer formats: multiple-choice (MC, $1{,}878$ questions) and Boolean (\emph{bool}, $1{,}122$ questions).
Our analysis surfaces two structural properties---\emph{heterogeneous corpora} and a \emph{long tail of compound temporal categories}---each of which directly shapes a component of the method.

\subsection{Corpus Composition}

The questions are sourced from four datasets with markedly different visual and audio characteristics (Table~\ref{tab:corpora}).
Each video's source is recoverable from its identifier prefix.
The corpora span a wide range of clip length and action density.
Among them, only CrossTask provides effective instructional narration; STAR, AGQA, and Breakfast are treated as having no reliable audio signal.

\begin{table}[b]
  \centering
  \resizebox{\linewidth}{!}{
  \begin{tabular}{llrrl}
    \toprule
    \textbf{Corpus} & \textbf{Type} & \textbf{\#Vid.} & \textbf{\#Q} & \textbf{Audio / length} \\
    \midrule
    CrossTask~\cite{zhukov2019crosstask} & instructional      & $472$ & $729$ & effective / long \\
    STAR~\cite{wu2021star}               & situated activity  & $466$ & $777$ & none / short \\
    AGQA~\cite{grundemclaughlin2021agqa} & activity QA        & $473$ & $760$ & none / short \\
    Breakfast~\cite{kuehne2014language}  & kitchen webcam     & $439$ & $734$ & none / multi-min \\
    \bottomrule
  \end{tabular}
  }
  \caption{Test-split composition by source corpus. Only CrossTask provides effective audio; the remaining corpora are treated as visual-only, with different clip lengths and action densities.}
  \label{tab:corpora}
\end{table}

This heterogeneity is the primary motivation for the \textbf{corpus-adaptive budgeting} in Section~\ref{sec:method}: a single global frame/step budget is mismatched to inputs that range from narrated CrossTask clips to silent, action-dense recordings.
The kitchen-webcam corpus in particular pairs high action density with no effective audio, so it benefits from a larger frame budget, finer temporal segmentation, and visual-only reasoning.

\subsection{Temporal-Category Distribution}

We infer each question's temporal-logic category from its surface template, mapping $2{,}821$ of the $3{,}000$ questions onto the benchmark's operator set; the remaining $179$ are spatial/object-relation MC questions that do not correspond to a temporal operator.
Table~\ref{tab:catdist} shows the resulting distribution.

\begin{table}[t]
  \centering
  \footnotesize
  \begin{tabular}{lrrr}
    \toprule
    \textbf{Category} & \textbf{Total} & \textbf{bool} & \textbf{MC} \\
    \midrule
    Until                & $185$ & $71$  & $114$ \\
    Since                & $112$ & $58$  & $54$  \\
    Disjoint             & $204$ & $108$ & $96$  \\
    Implies              & $142$ & $47$  & $95$  \\
    Before               & $161$ & $55$  & $106$ \\
    Next / After         & $155$ & $77$  & $78$  \\
    Co-Occur             & $161$ & $88$  & $73$  \\
    Immediate Next       & $157$ & $75$  & $82$  \\
    Always Before        & $132$ & $69$  & $63$  \\
    Always Next          & $136$ & $71$  & $65$  \\
    Always Co-Occur      & $34$  & $10$  & $24$  \\
    Strict A,B,C         & $166$ & $41$  & $125$ \\
    Loose A,B,C          & $549$ & $225$ & $324$ \\
    A always before B,C  & $527$ & $127$ & $400$ \\
    \midrule
    Other / non-temporal & $179$ & $0$   & $179$ \\
    \bottomrule
  \end{tabular}
  \caption{Inferred temporal-category distribution of the test split. Compound three-event categories (Strict/Loose A,B,C and A-always-before-B,C) dominate the mass, especially in MC format.}
  \label{tab:catdist}
\end{table}

The distribution is heavily long-tailed toward \textbf{compound, three-event categories}: \emph{Loose A,B,C} and \emph{A always before B,C} alone account for over a third of the split, and together with \emph{Strict A,B,C} they are strongly concentrated in MC format ($849$ MC questions).
These categories require reasoning over a full event chain $A \!\to\! B \!\to\! C$ rather than a single pairwise relation, and are the hardest to answer without an explicit timeline.
This observation motivates two method components: the \textbf{timeline-first protocol} for chain categories and the larger iteration/thinking budgets allocated to them.

\subsection{Summary of Design Implications}

The analysis yields a direct mapping from dataset structure to method:
(i) corpus heterogeneity $\Rightarrow$ corpus-adaptive sampling budgets and audio gating; and
(ii) the long tail of compound chain categories $\Rightarrow$ timeline-first reasoning with deeper iteration.
Section~\ref{sec:method} develops each of these components.

\section{Method}
\label{sec:method}

We cast temporal-logic VideoQA as an \emph{evidence-seeking} problem rather than a fixed-budget classification problem, similar in spirit to long-horizon video agents that seek evidence through tools~\cite{lin2026videoseek}.
Instead of feeding a vision–language model (VLM) a single uniformly sampled set of frames, our agent actively decides \emph{where} in the video to look, building an explicit action timeline before committing to an answer.
The agent follows a \textbf{Think--Act--Observe} loop, following the general ReAct-style pattern of interleaving reasoning with actions~\cite{yao2023react,lin2026videoseek}, equipped with a multi-granular sampling toolkit and a question-type-aware exploration policy.
The design is backbone-agnostic: the loop drives any frame-conditioned VLM through a uniform interface, so the same policy applies regardless of the specific model used.

\subsection{Overview}

\begin{figure*}[t]
  \centering
  \resizebox{0.96\textwidth}{!}{%
  \begin{tikzpicture}[
    node distance=9mm,
    io/.style={rectangle, rounded corners=5pt, draw=cvprblue, thick, fill=cvprblue!8, align=center, inner sep=4pt, font=\small, text width=26mm},
    proc/.style={rectangle, draw=black!55, thick, fill=black!3, align=center, inner sep=4pt, font=\small, text width=30mm},
    side/.style={rectangle, draw=black!55, thick, fill=black!3, align=center, inner sep=4pt, font=\small, text width=30mm},
    dec/.style={diamond, aspect=1.9, draw=black!55, thick, fill=orange!12, align=center, inner sep=1pt, font=\footnotesize},
    state/.style={rectangle, rounded corners=3pt, draw=black!40, dashed, fill=black!2, align=center, inner sep=4pt, font=\footnotesize, text width=44mm},
    arr/.style={-{Stealth[length=2mm]}, thick, draw=black!70},
    larr/.style={-{Stealth[length=2mm]}, thick, draw=cvprblue!85},
  ]
    \node[io] (q) {Video $V$ + Question $q$};
    \node[proc, right=of q] (cls) {\textbf{Classify} $q\to$ temporal category};
    \node[proc, right=of cls] (exp) {\textbf{Explore}: initial timestamped sampling};
    \node[proc, right=of exp] (think) {\textbf{Think}: reason over full history};
    \node[dec, right=of think] (dec) {commit\\answer?};
    \node[io, right=12mm of dec] (fin) {\textbf{Finalize} $\to$ \texttt{ANSWER} $\hat{a}$};
    \node[side, below=14mm of think] (obs) {\textbf{Observe}: new timestamped frames};
    \node[side, below=14mm of dec] (act) {\textbf{Act}: call tool \texttt{skim}/\texttt{focus}/\texttt{scan}};
    \node[state, above=9mm of think] (st) {\textbf{State}: timestamped observations + reasoning trace};

    \draw[arr] (q)--(cls);
    \draw[arr] (cls)--(exp);
    \draw[arr] (exp)--(think);
    \draw[arr] (think)--(dec);
    \draw[arr] (dec)-- node[above,font=\scriptsize]{yes} (fin);
    \draw[arr] (dec)-- node[right,font=\scriptsize]{no} (act);
    \draw[larr] (act)-- node[above,font=\scriptsize]{} (obs);
    \draw[larr] (obs)--(think);
    \draw[arr,<->,dashed,draw=black!50] (st)--(think);
  \end{tikzpicture}}
  \caption{Agent pipeline. After classification, an initial exploration produces the first timestamped frames; the agent then iterates the \textbf{Think--Act--Observe} loop (bottom), with each new observation accumulated in a persistent \emph{state} that is re-injected into the next prompt. The loop terminates---committing a final answer---once the VLM is confident and the minimum-step gate is satisfied, or when the step/frame budget is exhausted.}
  \label{fig:pipeline}
\end{figure*}
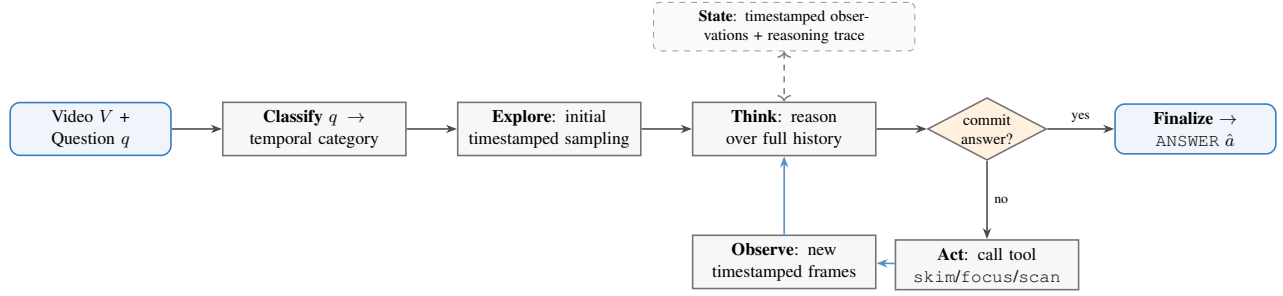

Given a video $V$, a question $q$, and (for multiple-choice questions) a set of options $\mathcal{C}$, the agent produces an answer $\hat{a}$ through an iterative loop (Figure~\ref{fig:pipeline}):
\begin{enumerate}
    \item \textbf{Classify.} $q$ is mapped to a temporal-logic category, which selects a tailored exploration strategy.
    \item \textbf{Explore.} An initial sampling tool produces a first set of timestamped frames.
    \item \textbf{Think--Act--Observe.} The VLM reasons over the accumulated observations and either (i) requests a more targeted sampling action, or (ii) commits to a final answer.
    \item \textbf{Finalize.} Once confident---or when the step/frame budget is exhausted---the VLM emits a parsed answer.
\end{enumerate}
A state object accumulates the timestamped observations and the per-step reasoning traces, which are re-injected into every subsequent prompt so that the VLM reasons over the \emph{full} exploration history rather than the latest frames alone.

\subsection{Multi-Granular Sampling Toolkit}

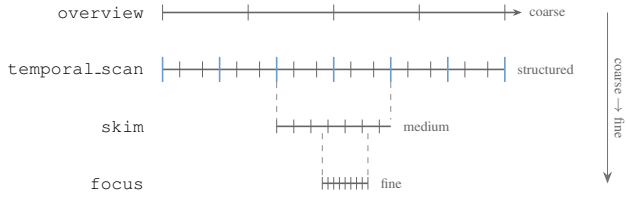
\begin{figure}[!t]
  \centering
  \resizebox{\linewidth}{!}{%
  \begin{tikzpicture}[x=1mm,y=1mm,
    strip/.style={thick, draw=black!60},
    tick/.style={draw=black!60},
    seg/.style={thick, draw=cvprblue!75},
    lbl/.style={font=\small\ttfamily, anchor=east},
    sub/.style={font=\scriptsize, anchor=west, text=black!60},
    zoom/.style={dashed, draw=black!45},
  ]
    \draw[strip,-{Stealth[length=1.6mm]}] (0,30)--(63,30) node[sub]{coarse};
    \foreach \x in {0,15,30,45,60} \draw[tick] (\x,28.6)--(\x,31.4);
    \node[lbl] at (-2,30) {overview};
    \draw[strip] (0,20)--(60,20);
    \foreach \s in {0,1,2,3,4,5} \draw[seg] ({\s*10},17.8)--({\s*10},22.2);
    \draw[seg] (60,17.8)--(60,22.2);
    \foreach \s in {0,1,2,3,4,5} {\draw[tick] ({\s*10+3},18.6)--({\s*10+3},21.4); \draw[tick] ({\s*10+7},18.6)--({\s*10+7},21.4);}
    \node[lbl] at (-2,20) {temporal\_scan};
    \node[sub] at (61,20) {structured};
    \draw[strip] (20,10)--(40,10);
    \foreach \x in {20,23,...,40} \draw[tick] (\x,8.8)--(\x,11.2);
    \node[lbl] at (-2,10) {skim};
    \node[sub] at (41,10) {medium};
    \draw[strip] (28,0)--(36,0);
    \foreach \x in {28,29,...,36} \draw[tick] (\x,-1.1)--(\x,1.1);
    \node[lbl] at (-2,0) {focus};
    \node[sub] at (37,0) {fine};
    \draw[zoom] (20,17.8)--(20,11.2); \draw[zoom] (40,17.8)--(40,11.2);
    \draw[zoom] (28,8.8)--(28,1.1); \draw[zoom] (36,8.8)--(36,1.1);
    \draw[-{Stealth[length=2mm]},draw=black!55] (78,30)--(78,0);
    \node[rotate=-90, anchor=south, font=\scriptsize, text=black!60] at (78,15) {coarse $\to$ fine};
  \end{tikzpicture}}
  \caption{Coarse-to-fine sampling toolkit on a shared time axis. \texttt{overview} samples the full clip sparsely; \texttt{temporal\_scan} divides it into $n$ labeled segments ($k$ frames each) for timeline construction; \texttt{skim} and \texttt{focus} progressively zoom into the window that determines the answer. Every returned frame is interleaved with its absolute timestamp.}
  \label{fig:toolkit}
\end{figure}

Temporal-logic questions vary widely in the temporal precision they demand: an \emph{eventually} question only needs to confirm an action occurs \emph{somewhere}, whereas an \emph{immediately-after} question hinges on the exact boundary between two adjacent actions.
To support both, we expose four tools spanning coarse-to-fine temporal granularity (Table~\ref{tab:tools}, Figure~\ref{fig:toolkit}).

All tools return frames \emph{interleaved with their wall-clock timestamps} (in seconds). This \textbf{frame--timestamp interleaving} is central to the method: it grounds every observation on an absolute time axis, allowing the VLM to express and verify temporal relations (before/after/overlap) as numerical comparisons rather than relying on implicit frame ordering.

\begin{table}[!b]
  \centering
  \resizebox{\linewidth}{!}{
  \begin{tabular}{llp{4.6cm}}
    \toprule
    \textbf{Tool} & \textbf{Granularity} & \textbf{Purpose} \\
    \midrule
    \texttt{overview} & coarse & Uniform global sampling for video structure. \\
    \texttt{temporal\_scan} & structured & Split the video into $n$ equal segments and sample $k$ frames per segment to build an action timeline. \\
    \texttt{skim} & medium & Medium-density sampling of a chosen $[t_s, t_e]$ window. \\
    \texttt{focus} & fine & Dense sampling of a narrow window for action-boundary analysis. \\
    \bottomrule
  \end{tabular}
  }
  \caption{Multi-granular toolkit. \texttt{temporal\_scan} provides structured coverage with explicit segment boundaries and is the default entry point for timeline construction.}
  \label{tab:tools}
\end{table}

The \texttt{temporal\_scan} tool is the workhorse for timeline construction.
By dividing $V$ into $n$ equal segments and sampling $k$ frames per segment, it provides denser, structurally labeled coverage than a flat uniform sample, so the VLM can localize each candidate action to a segment before zooming in with \texttt{skim}/\texttt{focus}.

\subsection{Question-Type-Aware Exploration}

A lightweight rule-based classifier maps $q$ to one of the benchmark's temporal categories (e.g.\ \emph{before}, \emph{after}, \emph{immediately-after}, \emph{until}, \emph{since}, \emph{overlap}, \emph{not-overlap}, \emph{imply}, three-event \emph{chains}, etc.) via ordered keyword/structure patterns, falling back to a generic strategy when no pattern matches.
Each category is associated with an \textbf{exploration strategy} that specializes the loop along several axes:
\begin{itemize}
    \item \textbf{Initial tool} (\texttt{temporal\_scan} vs.\ \texttt{overview}) and initial frame budget.
    \item \textbf{Preferred follow-up tool} (\texttt{focus} for boundary-sensitive types such as \emph{immediately-after}; \texttt{skim} for ordering types such as \emph{before}/\emph{after} and chains).
    \item \textbf{Iteration depth}: the maximum number of steps and whether iteration is needed at all (simple existence checks such as \emph{eventually}/\emph{throughout} may answer in a single pass).
    \item \textbf{A type-specific reasoning prompt} that encodes the formal semantics of the operator (e.g.\ ``\emph{A before B} means $A$'s last occurrence precedes $B$'s first occurrence'') together with common pitfalls (e.g.\ confusing a continuous \emph{state} such as \emph{holding} with a \emph{point} action such as \emph{picking up}).
    \item \textbf{A thinking-token budget} scaled to category difficulty, larger for multi-event chains and smaller for simple existence checks.
\end{itemize}
For the hardest three-event chain categories, the prompt additionally enforces a \textbf{timeline-first} protocol: the VLM must emit an explicit \texttt{Timeline: [(action, $t_s$, $t_e$), \dots]} before it is permitted to select an answer.

\subsection{The Think--Act--Observe Loop}

After the initial exploration, the agent alternates between observation and action.
At each step the VLM receives the new timestamped frames, the toolkit description, all prior reasoning, and the type-specific hint, then responds with either a tool call (parsed from a JSON block or an \texttt{ACTION: tool(args)} form) or a final answer (\texttt{ANSWER: $\langle$letter/Yes/No$\rangle$}).

\paragraph{Minimum-step gating.}
A recurring failure mode of single-pass prompting is \emph{premature commitment}: the VLM writes an answer on the first step without ever verifying its tentative timeline.
For ordering and chain categories on long videos, we therefore impose a \emph{minimum} number of exploration steps: any early \texttt{ANSWER} is overridden and replaced by a forced follow-up tool call until the minimum is met.
Short videos bypass this gate, since dense sampling already covers them completely in one pass (below).

\paragraph{Adaptive frame and step budgets.}
Budgets are adapted to the input rather than fixed globally.
\emph{Short videos} (below a per-category duration threshold) are handled by \textbf{dense uniform sampling} of the full clip in a single step, which is both cheaper and more reliable than iterative seeking when the entire video fits in the budget.
\emph{Longer videos} use the iterative loop with \texttt{temporal\_scan} initialization.
We further apply \textbf{corpus-adaptive} budgeting: clips drawn from sources with denser, more repetitive low-level actions and no effective audio receive a higher minimum frame budget, finer \texttt{temporal\_scan} segmentation, and a higher minimum step count, because fine action-boundary discrimination is the bottleneck there; CrossTask clips can additionally use their instructional narration.

\subsection{Auxiliary Speech Context}

For CrossTask videos, we attach an ASR transcript as additional temporal evidence: timestamped transcript segments are prepended to the question so the VLM can align spoken narration with visual events.
For STAR, AGQA, and Breakfast, the transcript is omitted because these corpora do not provide effective audio for our setting.

\section{Experiments}
\label{sec:experiments}

This section summarizes the experimental setup and official final test-set result.
The system is used strictly as an inference-time agent: no task-specific training, fine-tuning, or validation-set model selection is performed.
We therefore report only the final official test performance rather than training curves or ablation studies.

\subsection{Experimental Setup}

We use \textbf{Gemini 3.1 Pro}~\cite{geminiteam2026gemini31pro} as the VLM backbone for all submitted predictions.
The model is queried through the evidence-seeking agent described in Section~\ref{sec:method}: each question is first mapped to a temporal category, the corresponding sampling policy selects timestamped video frames, and the accumulated observations are passed to Gemini 3.1 Pro for reasoning and answer generation.
All task adaptation is performed at inference time through the question classifier, prompt templates, sampling schedules, and tool-calling loop; no model weights are updated.

The inference configuration fixes the backbone model, maximum frames per question, maximum agent steps, category-specific minimum steps, corpus-specific budget multipliers, ASR usage, and random seed.
For each example, the system records the predicted temporal category, sampled timestamps, tool calls, intermediate reasoning summaries, raw model output, and parsed final answer.
The leaderboard submission contains the compact answer file parsed from the model outputs.

\subsection{Evaluation Protocol}

We evaluate on the official TimeLogic test split using the challenge's \textbf{AvgAcc} metric.
For multiple-choice questions, the parser extracts a single option letter from the model output; for Boolean questions, it extracts \texttt{Yes} or \texttt{No}.
Invalid or unparsable outputs are counted as incorrect.
All reported numbers correspond to the final submitted inference run with fixed sampling parameters and random seed.
Our final submission obtains an official test-set \textbf{AvgAcc of 77.13}.
This is the only quantitative result reported in this challenge submission.

\section{Conclusion}
\label{sec:conclusion}

We presented an evidence-seeking agent for the TimeLogic Challenge that converts temporal-logic VideoQA from a single-pass frame classification problem into an iterative search for timestamped evidence.
The method combines benchmark-driven question categorization, multi-granular sampling, explicit timestamp interleaving, category-specific reasoning prompts, and corpus-adaptive budgets.
This design requires no task-specific training and can be paired with different frame-conditioned VLM backbones.

Our analysis suggests that temporal grounding is the key bottleneck: strong VLMs can often recognize the relevant actions once they are shown the right frames, but they need an explicit mechanism for finding, ordering, and verifying those frames.
The final system applies this mechanism with Gemini 3.1 Pro as the VLM backbone and achieves an official test-set AvgAcc of 77.13.

{
    \small
    \bibliographystyle{ieeenat_fullname}
    \bibliography{main}
}


\end{document}